\documentclass{ws-p8-50x6-00}

\def\Gev{{\hbox {\rm GeV}}}

\begin{document}

\title{MRST Global Fit Update}

\author{R.S. Thorne}

\address{Cavendish Laboratory, University of Cambridge\footnote{Royal 
Society University Research Fellow}, Madingley Road,\\
Cambridge, CB3 0HE, U.K.}

\author{A.D. Martin and W.J. Stirling}

\address{Department of Physics, University of Durham, Durham, DH1 3LE, U.K.}

\author{R.G. Roberts}

\address{Rutherford Appleton Laboratory, Chilton, Didcot, Oxon, OX11 0QX, 
U.K.}  


\maketitle

\abstracts{
We discuss the impact of the most recent data on the MRST global analysis  
- in particular the new high-$E_T$ jet data and their implications 
for the gluon and the new small $x$ structure function data.
In the light of these new data we also consider the uncertainty in predictions
for physical quantities depending on parton distributions, concentrating on 
the $W$ cross-section at hadron colliders.}

There has recently been a great deal of updated data which help to determine 
parton distributions, particularly on small $x$ structure functions from
HERA\cite{H1,ZEUS} and on inclusive jets from the Tevatron\cite{D0,CDF}.
In both cases these new data are both more precise and extend the kinematic 
range, and thus determine the parton distributions more accurately than ever
before. The MRST global fit is performed by minimizing the $\chi^2$ 
for all data
adding statistical and systematic errors in quadrature  
(this procedure will be improved before 
producing finalized distributions), except for the jet data where the 
dominant correlated errors render a proper treatment absolutely essential. 
The evolution begins at $Q_0^2=1\Gev^2$, and data are cut for 
$Q^2 < 2 \Gev^2$ and $W^2 < 10 \Gev^2$ in order to exclude regions where 
higher twist and/or higher orders in $\alpha_S$ are expected to play an 
important role.  

The main effect of the new data is to constrain the gluon distribution, and
hence, via evolution the small $x$ quark distributions. The jet data are 
important in determining the high $x$ gluon. Previously\cite{MRST98,MRST99} 
MRST 
used prompt photon data to perform this role, but it has become clear that 
there are both large theoretical uncertainties and conflicts between 
data sets, and we now drop this approach.
The best fit to jet data produces a high $x$ gluon in between our
previous $g$ and $g \uparrow$ distributions, i.e. $xg(x) \sim (1-x)^{5.5}$. 
If the gluon becomes smaller at high $x$ the fit to the jets becomes 
unacceptable. If it becomes larger, then although the jet data are initially
supportive, the fit to DIS data deteriorates dramatically. Hence, there is
an uncertainty of $\sim 20 \%$ at $Q^2 = 20 \Gev^2$, $x=0.4$. 

\begin{figure}
\centerline{\hspace{-0.54in}\epsfig{figure=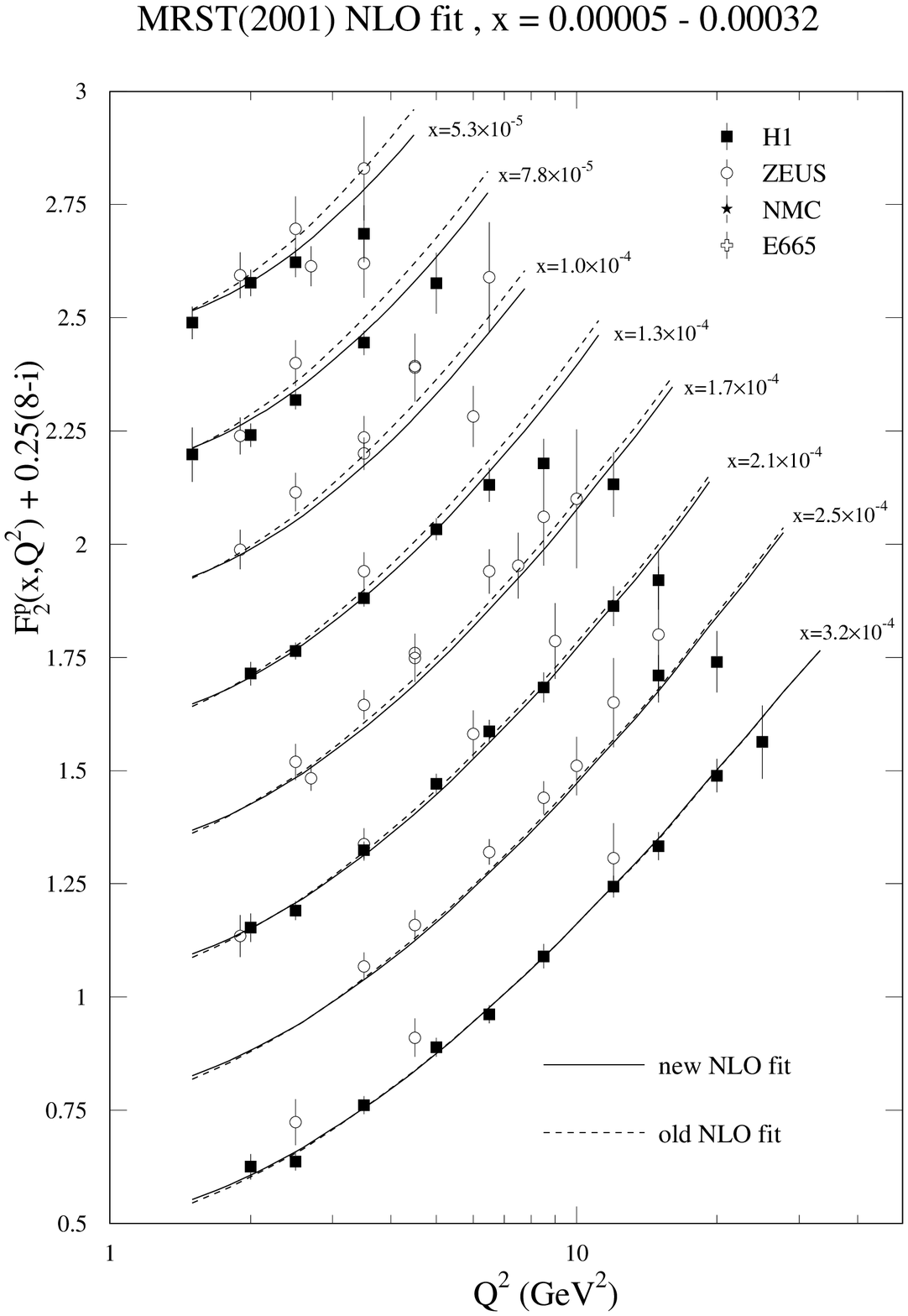,height=3.3in,width=2.3in}
\epsfig{figure=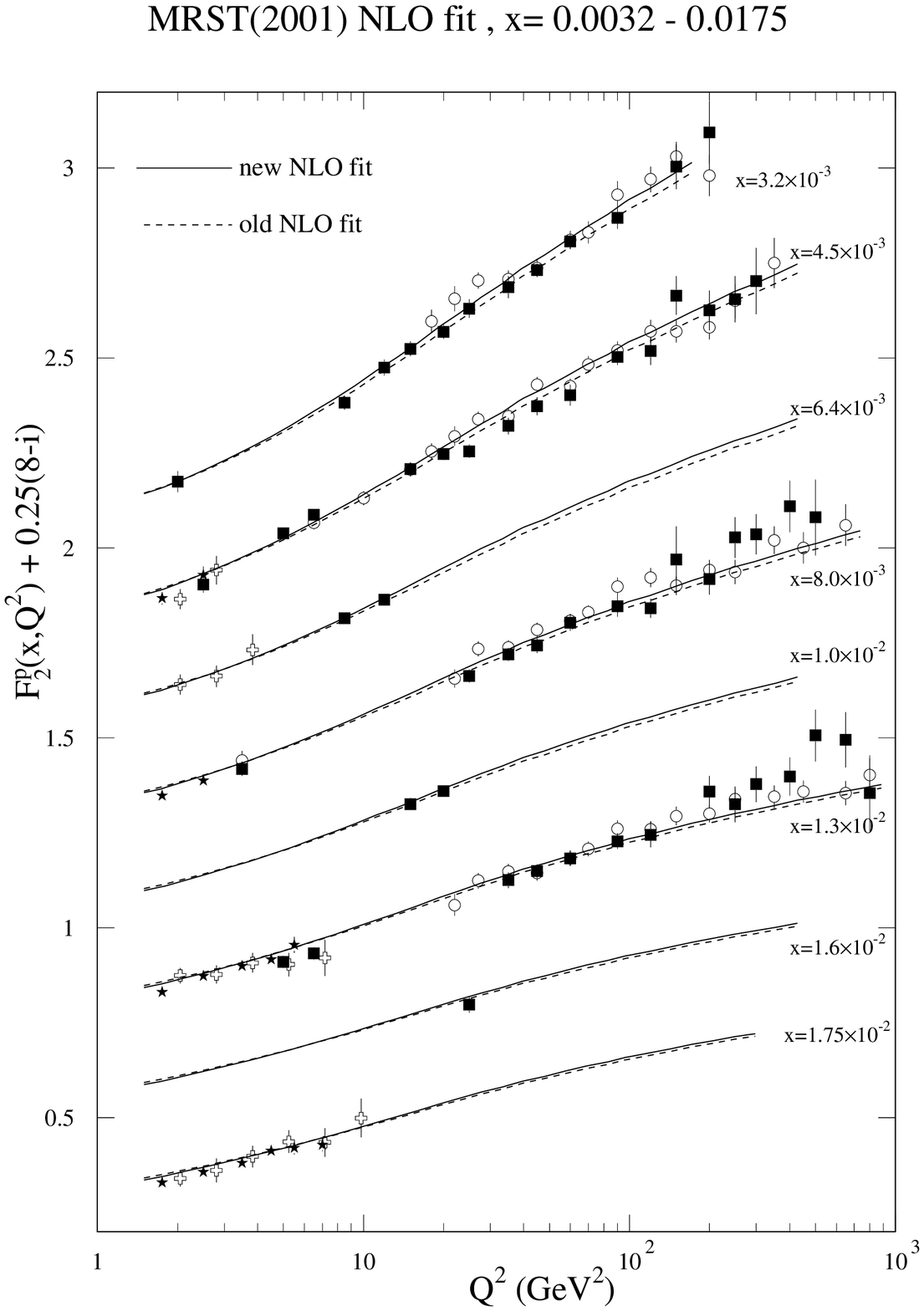,height=3.3in,width=2.3in}}
\vspace{-0.4in}
\caption{Comparison of the MRST2001 structure function at small $x$ to that 
from MRST 1999 along with the new data from H1 and ZEUS.}
\vspace{-0.2in}
\label{fig:mrst2001}
\end{figure}

The improved HERA data also significantly affect the gluon. 
In order to obtain an acceptable fit the parameterization at $Q_0^2$ 
is extended to the form 
\begin{equation}
xg(x,Q_0^2)=A_g(1-x)^{\eta_g}(1+\epsilon_gx^{0.5}+\gamma_g x)x^{\delta_g}
-A_-(1-x)^{\eta_-}x^{-\delta_-},
\label{eq:gluon}
\end{equation}            
which allows the gluon to become negative at small $x$. $\eta_-$ is 
fixed at $\sim 10$ so that the large $x$ form is unchanged, and
$\delta_- \sim 0.2$ - the gluon becoming quite negative at the smallest $x$.
It becomes positive for all $x>10^{-5}$ for $Q^2 > 2-3 \Gev^2$.  
A good fit is obtained for the HERA structure function data as seen in fig. 1.
Compared to the last MRST fit $F_2(x,Q^2)$ is flatter in $Q^2$ at the 
smallest $x$, but steeper at $0.05 < x < 0.0004$. In this intermediate range 
the data would prefer an even higher $dF_2(x,Q^2)/d\ln Q^2$. 
Also, we find that the jet and DIS data push in opposite directions for 
$\alpha_S(M_Z^2)$. Once the high $x$ constraint on the gluon is imposed, the
DIS data prefer $\alpha_S(M_Z^2)=0.121$ while the jet data prefer 
$\alpha_S(M_Z^2)=0.117$, a compromise of $0.119$ being reached.

In order to improve the analysis of physical quantities sensitive to structure 
functions and make truly quantitative predictions there are a number of 
issues to address. Many important ones are theoretical, such as
higher orders, resummation of $\ln(1-x)$ and $\ln(1/x)$ terms, higher twist, 
etc.. However, a direct issue which has been a recent focus of attention 
is the uncertainty 
due to the experimental errors on current data. Rather than obtain parton 
distributions with errors\cite{Botje,CTEQH,Gielea}, 
we follow 
our original suggestion\cite{MRST99} and look at the error 
on a physical quantity determined by the parton 
distributions\cite{CTEQL,Gieleb}, in practice the 
$W$ cross-section at hadron colliders.   
     
We study $\sigma_W$ for the Tevatron and LHC, which probes mainly the 
quarks in the region $0.5 < x < 0.005$ for the former and $0.5 <x< 0.00007$
for the latter. For our best fit we find
\begin{equation}
\sigma_W({\rm Tev}) = 22.3nb, \qquad \sigma_W({\rm LHC}) = 192nb.
\label{eq:sigma}
\end{equation}  
In our previous study we examined the variation in $\sigma_W$ due to 
variation in normalization of data, $\alpha_S(M_Z^2)$, form of high $x$ gluon
etc, rather than the uncertainty on the data. In order to do this the best 
way to proceed is to perform the global fit whilst constraining the value of 
the quantity in question, i.e. to use the Lagrange multiplier 
method\cite{CTEQL}. However, this previous study\cite{CTEQL} produced a 
large value for the 
uncertainty in $\sigma_W$ of order $5-8\%$. We have now performed an 
analogous study using the most up-to-date data, and also using 
a different method of determining the limits of an unacceptable fit. 
We impose the rough criterion that no data set has a less than $1\%$ 
confidence level. This leads to an estimated uncertainty on $\sigma_W$ of 
about $\pm 2\%$. In short, for the upper limit on $\sigma_W({\rm LHC})$
the fit to H1 data fails, for the lower limit on $\sigma_W({\rm LHC})$
the fit to ZEUS data fails, for the upper limit on $\sigma_W({\rm Tev})$
the fit to NMC proton data fails and for the lower limit on 
$\sigma_W({\rm Tev})$ the fit to NMC $n/p$ data fails. Often at least one 
other data set has a deteriorating quality of 
fit and the increase in global $\chi^2$ is
about 80. A roughly symmetric deterioration in the fit quality to CCFR,
for example, is shown in fig. 2. 
\begin{figure}
\centerline{\hspace{-0.54in}
\epsfig{figure=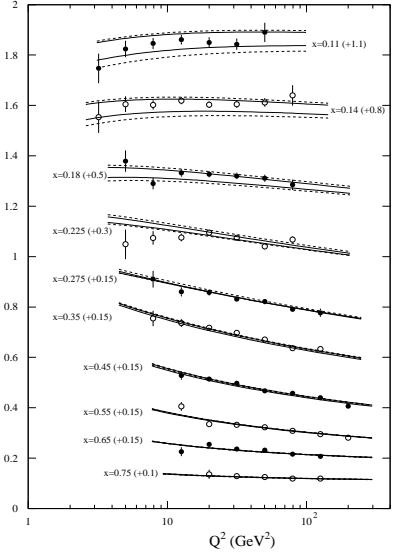,height=3.3in,width=2.3in}}
\vspace{-0.4in}
\caption{Comparison of CCFR $F_3(x,Q^2)$ data to theory for 
$\sigma_W({\rm Tev})$ changing in magnitude by factors of $1.021$, $0.973$,
$1.034$ and $0.956$.}
\vspace{-0.2in}
\label{fig:ccfr}
\end{figure}
   
We conclude that the uncertainty in $\sigma_W$, and similarly $\sigma_Z$,
due to experimental 
errors is rather small. We can repeat the procedure for a variety of other
observables, e.g. Higgs production will be directly sensitive to the small
$x$ gluon rather than small $x$ quarks, and the uncertainty due to experimental
errors will obviously be largest for quantities sensitive to the large $x$ 
gluon. However, we also need to consider the uncertainty due to theoretical 
errors. As was shown in \cite{MRSTNNLO}, the perturbative series seems fairly
convergent for quark sensitive processes (though an improvement is perhaps 
needed in $dF_2(x,Q^2)/d\ln Q^2$) probably because it is the quarks that are 
directly tied down by experiment at present. However, the expansion 
is less reliable for 
gluon sensitive quantities - especially in the region where resummations are 
expected to be important. Hence, although we will continue detailed work into 
uncertainties due to experimental errors, we believe that the great precision 
of data now leads to the theoretical errors being dominant in most cases, and
suggest that this should be a major area of study in the immediate future.


\end{document}